\documentclass[twocolumn]{aastex62}
\usepackage{aas_macros}
\usepackage{rotating}
\usepackage{comment}
\usepackage{float} \usepackage{amsmath,amssymb}
\usepackage{natbib} % For use with bibtex
\usepackage{graphicx} % For included graphics
\usepackage{color} \usepackage{verbatim}
\usepackage{rotating}
\usepackage{CJK}
\usepackage{amsmath}
\usepackage{bm}
\usepackage{physics}
\usepackage[T1]{fontenc}

\newcommand{\cntextsc}[1]{\begin{CJK*}{UTF8}{gbsn}#1\end{CJK*}}
\newcommand{\jptext}[1]{\begin{CJK}{UTF8}{min}#1\end{CJK}}
\newcommand{\proptosim}{\mathrel{\vcenter{
 \offinterlineskip\halign{\hfil$##$\cr
 \propto\cr\noalign{\kern2pt}\sim\cr\noalign{\kern-2pt}}}}}

\newcommand{\unit}[1]{{\rm\, #1}}

\newcommand{\mean}[1]{\langle #1\rangle}

\renewcommand{\min}{\mathrm{min}}
\renewcommand{\max}{\mathrm{max}}
\newcommand{\au}{\textsc{au}}
\newcommand{\cm}{\unit{cm}}

\newcommand{\g}{\unit{g}}

\newcommand{\K}{\unit{K}}

\renewcommand{\micron}{\unit{\mu m}}

\newcommand{\eV}{\unit{eV}}
\newcommand{\keV}{\unit{keV}}
\newcommand{\s}{\mathrm{s}}
\newcommand{\yr}{\mathrm{yr}}

\newcommand{\ang}{\ensuremath{\mathrm{\AA}}}

\newcommand{\am}{\mathrm{Am}}

\renewcommand{\d}{\mathrm{d}}

\newcommand{\Gr}{\ensuremath{\mathrm{gr}}}
\newcommand{\gr}{\ensuremath{\mathrm{gr}}}
\newcommand{\acc}{\mathrm{acc}}

\newcommand{\wind}{\mathrm{wind}}

\newcommand{\wb}{\mathrm{wb}}
\newcommand{\disk}{\mathrm{disk}}

\renewcommand{\O}{\mathrm{O}}     % Ohmic
\newcommand{\A}{\mathrm{A}}     % Ambipolar
\renewcommand{\H}{\mathrm{H}}     % Hall
\renewcommand{\P}{\mathrm{P}}     % P
\newcommand{\sigperH}{$\sigma_\Gr/\H$}
\newcommand*\chem[1]{\ensuremath{\mathrm{#1}}}
\newcommand{\pos}[1]{\ensuremath{\mathrm{#1}^+}}

\begin{document}
\begin{CJK*}{UTF8}{gkai}

\title{A Tale of Two Grains: impact of grain size 
  on ring formation via nonideal MHD processes}

\author{Xiao Hu (\cntextsc{胡晓})}
\affiliation{Department of Physics and Astronomy, University
  of Nevada, Las Vegas, 4505 South Maryland Parkway, Las
  Vegas, NV 89154} 
\author{Lile Wang (\cntextsc{王力乐})}
\affiliation{Center for Computational Astrophysics, Flatiron
  Institute, New York, NY 10010;
  lwang@flatironinstitute.org} 
\author{Satoshi Okuzumi (\jptext{奥住聡})}
\affiliation{Department of Earth and Planetary Sciences,
  Tokyo Institute of Technology, Meguro, Tokyo, 152-8551,
  Japan} 
\author{Zhaohuan Zhu (\cntextsc{朱照寰})}
\affil{Department of Physics and Astronomy, University of
  Nevada, Las Vegas, 4505 South Maryland Parkway, Las Vegas,
  NV 89154} 

\begin{abstract}
  Substructures in PPDs, whose
  ubiquity was unveiled by recent ALMA observations, are
  widely discussed regarding their possible origins. We
  carry out global full magnetohydrodynamic (MHD)
  simulations in axisymmetry, coupled with self-consistent ray-tracing
  radiative transfer, thermochemistry, and
  non-ideal MHD diffusivities. The abundance profiles of
  grains are also calculated based on the global dust evolution calculation,
  including sintering effects. We found that dust size plays a crucial role 
  in the ring formation around the snowlines of
  protoplanetary disks (PPDs) through the accretion process. Disk ionization structures
  and thus tensorial conductivities depend on the size of
  grains. When grains are significantly larger than PAHs,
  the non-ideal MHD conductivities change
  dramatically across each snow line of major volatiles, leading to a sudden change of
  the accretion process across the snow lines and the subsequent formation of gaseous rings/gaps there. 
  On the other hand, the variations of conductivities
  are a lot less with only PAH sized grains in disks and then these disks retain smoother radial
  density profiles across snow lines.
\end{abstract}


\section{Introduction}
\label{sec:intro}

Studying the dynamics of protoplanetary disks is crucial for not only constructing comprehensive planet
formation theory, but also  understanding the fine features
in recent high resolution observations. The {\it 2014 ALMA
  campaign} provided unprecedented details of the HL Tau
disk \citep{2015ApJ...808L...3A}. Patterns that are broadly
axisymmetric, such as multiple bright and dark rings in dust
continuum emission, are found to be common in protoplanetary
disks in subsequent observations \citep[e.g.][]
{2018ApJ...869L..42H, 2018ApJ...869...17L}.

Recent MHD simulations suggest that substructure formation
can be achieved by non-ideal MHD processes, including
Ohmic resistivity and ambipolar diffusion. Processes like redistribution of magnetic flux, direct feeding of
avalanche accretion, and midplane magnetic reconnection
\citep{2017MNRAS.468.3850S, 2018MNRAS.477.1239S,
  2019MNRAS.484..107S} can alter disk surface density. Meanwhile, snow lines modify dust drift, growth, and fragmentation, causing surface density variation radially \citep{2015ApJ...806L...7Z,2016ApJ...821...82O}. The abundance of dust is an important factor for disk's ionization, affecting the coupling between gas and magnetic fields \citep[e.g.,][]{2007Ap&SS.311...35W,2011ApJ...739...51B}. Near snow lines, we can expect the feedback from dust to non-ideal MHD, then to disk accretion itself. In \citet{2019arXiv190408899H}, this idea was tested by
incorporating snow line induced dust distribution and ionization change into MHD
disk simulations. The ionization structure was
pre-calculated with dust distribution from global dust
evolution including sintering effects
\citep{2016ApJ...821...82O}. The ionization structure and the non-ideal MHD
diffusivities have sharp changes at the snow lines due to the change of dust size there.
This leads to a discontinuous accretion flow across the snow lines. With time, such a discontinuous accretion flow
naturally produces 
gaps and adjacent rings at \chem{CO_2} and \chem{C_2H_6} snow lines. This
dust-to-gas feedback requires much less dust to take effect,
compared to hydrodynamic drag mechanism \citep[e.g.,][]{2017MNRAS.467.1984G}. 

However, the simple treatment of dust on magnetic diffusion is also
one of the major caveats of current non-ideal MHD
simulations. Grain-modified magnetic diffusivities in
protoplanetary disks were first explored in the very small
grain limit ($\sim$ nm size, or PAH scale). They are known
for substantially reducing the level of ionization, killing
magnetorotational instability
(MRI)\citep{2011ApJ...739...51B,2018arXiv181012330W}. Charged
grains, when very small, can be regarded as heavy ions in
diffusion calculations, while its geometric size starts to take
effect in larger ( sub-micron to micron-sized)
grains. \citet{2016ApJ...819...68X} systematically explored
the dependence of magnetic diffusivities on magnetic field
strength with various grain sizes, and found the dependence
of magnetic diffusion onto field strength shifts over grain
sizes.

Another major caveat of most global MHD simulations is the uncertain
disk temperature at the disk surface. Usually, a fixed
or fast relaxation temperature profile is employed, and the
initial density structure is derived based on hydrostatic
equilibrium in $R$-$z$ plane. The transition between the hot
atmosphere (also referred to as corona) and the cold, dense
disk is either sharp
\citep[e.g.,][]{2017MNRAS.468.3850S,2018MNRAS.477.1239S,
  2019MNRAS.484..107S} or with some artificial smooth
profile \citep[e.g.,][]{2019arXiv190408899H,
  2017A&A...600A..75B, 2017ApJ...836...46B}.

In this work, we aim to overcome these two shortcomings by
utilizing non-ideal MHD simulation with self-consistent
thermochemistry and radiative transfer \citep{2017ApJ...847...11W,
  2018arXiv181012330W}. The dust structure from global dust
evolution is input into the simulations, while Ohmic
resistivity and ambipolar diffusion profile can be obtained
in real-time based on the ionization structure from the thermochemistry calculations. 
This letter is organized as
follows. \S\ref{sec:grains} briefly described the global
dust evolution model and discussed the different roles
between small and large grains in magnetic
diffusion. \S\ref{sec:sim} summarizes the numerical methods
and parameter choices for our
simulations. \S\ref{sec:sim-results} presents diagnostics of
disk radial and vertical structures, focusing on relating
chemistry to ionization and at last accretion. We summarize
the paper in \S\ref{sec:summary}.

\section{Dust Grains and Diffusion of Fields}
\label{sec:grains}

Similar to \citet{2019arXiv190408899H}, we use the radial
profiles of dust surface density and particle size derived from a
one-dimensional global dust evolution model having multiple
snow lines by \citet{2019ApJ...878..132O}. In this model, the grain size distribution 
follows a power-law with minimum and
maximum grain sizes $a_{\gr,\min}$ and $a_{\gr,\max}$, respectively. The
power law is such that the vertically integrated number density of grains per unit grain radius $a_\Gr$ is proportional to $a_\Gr^{-3.5}$. We
fixed the minimum grain size $a_{\Gr,\min}$ as $0.1~\micron$,
whereas the maximum size $a_{\Gr,\max}$ and total dust surface
density $\Sigma_d$ are evolved by computing radial drift and
collisional growth/fragmentation. The calculation takes into
account the low stickiness of CO$_2$ ice
\citep{2019ApJ...878..132O} and aggregate sintering
\citep{2016ApJ...821...82O}.  In this 1-D dust evolutionary model, 
the fragmentation threshold velocity in regions where aggregate
sintering takes place is chosen
to be $40\%$ of the threshold for unsintered aggregates. We
assume a disk of weak gas turbulence with velocity
dispersion of 1.7\% of the sound speed. The gas
surface density distribution for the dust evolution
calculation is fixed in time and assumed to scale as
$R^{-1}$, where $R$ is the distance from the central star,
at $R < 150~\au$. We evolve the dust disk for 1.3 Myr, until the computed total millimeter fluxes from the dust
disk match the ALMA observations of the HL Tau disk
\citep{2015ApJ...808L...3A}.

\begin{figure}
    \centering
    \includegraphics[width=0.45\textwidth]{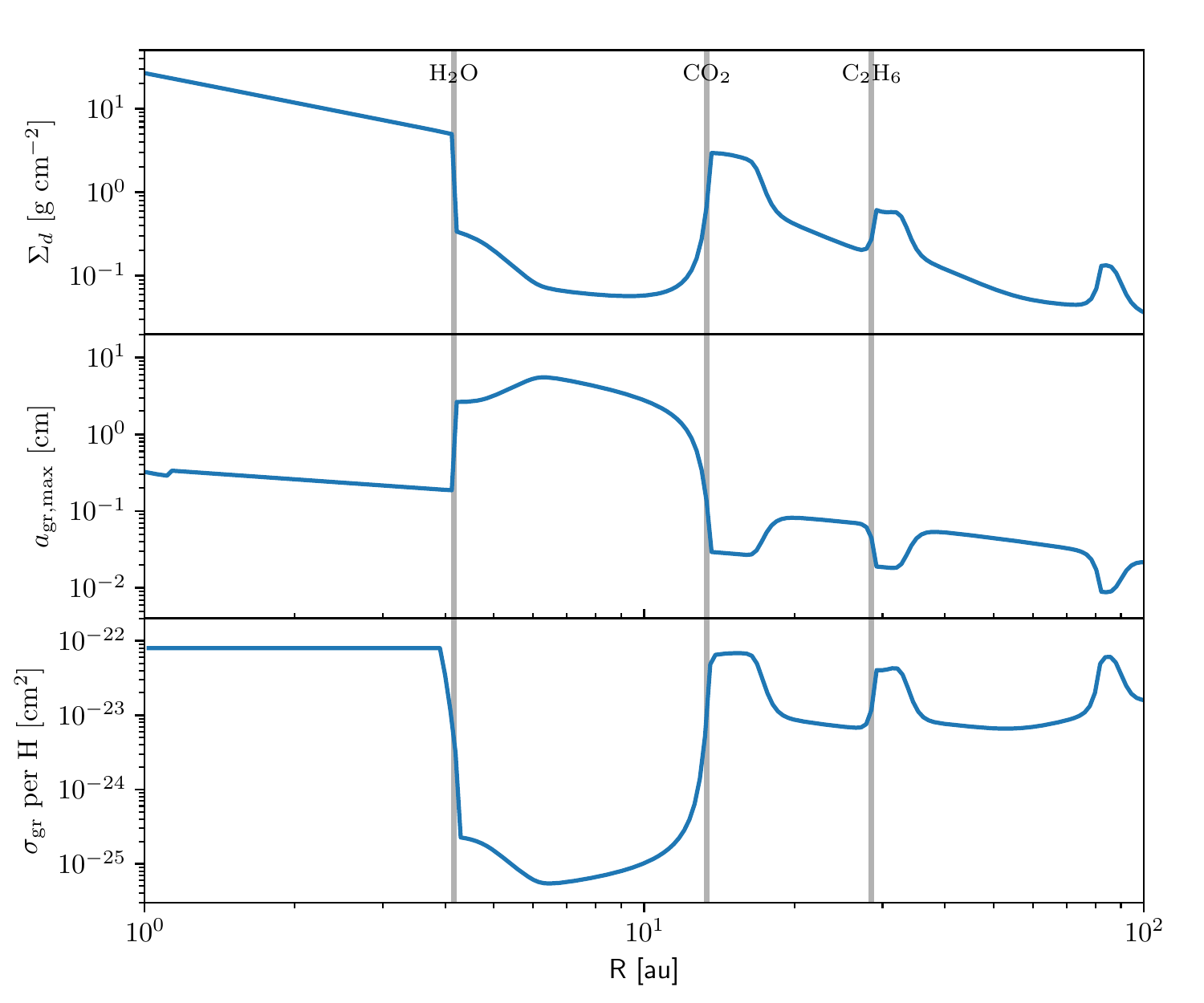}
    \caption{The upper and middle panels show the total dust
      surface density and maximum grain size, respectively,
      as a function of the distance $R$ from the central
      star. The lower panel shows dust's total geometric
      cross-section per Hydrogen nucleus. Gray vertical lines indicate location of three snow lines. Note the traffic jam interior to the H2O snow line is caused by the low stickiness of silicates, unrelated to sintering. The H$_2$O sintering zone corresponds to the small hill (at $R =$4--6 $\au$) at the bottom of the large gap in the $\Sigma_d$ profile.
      }
    \label{fig:dust_sig}
\end{figure}

The profile of $\Sigma_d$ 
is plotted in the upper panel of
Figure~\ref{fig:dust_sig} (also shown in the left, second
bottom panel in Figure 3 of
\citealt{2019ApJ...878..132O}). Because of sintering, dust
particles behind the snowlines CO$_2$, C$_2$H$_6$, and
CH$_4$ (located at $R = 13, 29, {\rm\ and\ } 82~\au$,
respectively) experience enhanced collisional fragmentation
and pile up there. Collisional fragmentation is also
enhanced interior to the \chem{H_2O} snow line (located at
$4~\au$) due to lack of water ice and results in another
traffic jam of dust in this region.

The adopted dust's  total  geometric  cross-section for chemistry calculation is
plotted in the lower panel of
Figure~\ref{fig:dust_sig}. Multiplying average cross-section
of grains of all sizes and total dust abundance (number
density over Hydrogen nucleus) we get dust's total geometric
cross-section per Hydrogen nucleus. This number quantifies
the contribution of dust particles to the whole chemical
network. We make it unchanged in different models that are
to be presented below. Note that we enhanced \sigperH ~ when $R<4~\au$, to ensure a moderate ionization fraction at inner disk. 

In a weakly ionized protoplanetary disk, the equation of
motion for charged species (whose inertia is usually
negligible, see e.g. \citealt{2016ApJ...819...68X}) is set
by the equilibrium between the Lorentz force and the
collision with neutrals,
\begin{equation}
  Z_je\left(\bm{E}^{'}
    + \frac{\bm{v}_j}{c}\times \bm{B}\right)
   = \gamma_j \rho m_j \bm{v}_j
\end{equation}
where for the charged species indicated by $j$, $Z_j$ is its
charge number (in unit charge $e$), $m_j$ is its mass,
$\bm{v}_j$ is its drift velocity relative to the neutral
background, 
$\gamma_j \equiv \mean{\sigma v}/(m+m_j)$ ($m$ is the
averaged particle mass of the neutrals) and $\mean{\sigma v_j}$ is its rate coefficient of
momentum transfer with neutrals.  The electric field
in the frame comoving with the neutrals is denoted by
$\bm{E}^{'}$.

In the case of very tiny grains, the interaction between
charged grains and neutral molecules are mediated by electric field
from induced electric dipoles in the neutrals. This $r^{-4}$
electric potential makes the collision rate coefficient
$\langle\sigma v\rangle_j$ independent of
temperature\citep[e.g.,][]{2011piim.book.....D}. When grains
get larger, their geometric cross section dominates their
interactions with neutrals. This transition brought by the
grain size is reflected by the Hall parameter, which is the
ratio between the gyrofrequency of charged species under
Lorentz force and their collision frequency with the
neutrals:
\begin{equation}
    \beta_j=\frac{\abs{Z_j}e B}{m_j c}\frac{1}{\gamma_j \rho}
    \label{eq:beta_j}
\end{equation}
In our simulation, we adopt the recipes in \citet
{2011ApJ...739...50B, 2014ApJ...791...72B} to calculate
collision coefficients between charged grains and neutrals:
\begin{equation}
\begin{split}
  \langle\sigma v\rangle_\Gr = \rm{max}&\bigg[1.3\times10^{-9}\abs{Z_\Gr},\\
  4\times10^{-3}&\left(\frac{a_\gr}{1\rm{\mu
        m}}\right)^{2}\left(\frac{T}{100\rm{K}}\right)^{1/2}\bigg]\rm{cm^3s^{-1}}
    \end{split}
    \label{eq:sig_v}
\end{equation}
So the transition from electric potential dominated cross section to geometric cross section is at $\sim nm$ scale. Given $T$=100K, any single charged grain with $a_\Gr > 5.7\times10^{-8}~\cm$ needs to consider geometric effect when calculating collision coefficient. This difference
is shown in the Ohmic, Hall and Pederson conductivities:
\begin{equation}
\begin{split}
  & \sigma_\O=\frac{ec}{B}\sum_j n_j|Z_j|\beta_j\ ,
  \\
  & \sigma_\H=\frac{ec}{B}\sum_j\frac{n_j Z_j}{1+\beta_j^2}
  \ ,\\
  & \sigma_\P=\frac{ec}{B}\sum_j\frac{n_j
    |Z_j|\beta_j}{1+\beta_j^2}\ .
\end{split}
\label{eq:sigfull}
\end{equation}
Here the summation index $j$ runs through all charged
species, with $n_j$ indicating the number density and charge
of individual charged species. Since
$\beta_{gr},\beta_i \ll 1$ stands true throughout most
regions of our disk, the difference between ions and grains
is negligible for Hall conductivity $\sigma_\H$. For Ohmic
and Pederson conductivities, the contribution of larger grains
can be orders of magnitude smaller than very small grains
and ions. Using these conductivities, the general expressions for the
three magnetic diffusivities are \citep{2011ApJ...739...50B,
  2018arXiv181012330W}.
\begin{equation}
\begin{split}
  &\eta_\O=\frac{c^2}{4\pi}\frac{1}{\sigma_\O}\ ,\quad
  \eta_\H=\frac{c^2}{4\pi}
  \frac{\sigma_H}{\sigma_\H^2+\sigma_\P^2}\ ,\\
  &
  \eta_\A=\frac{c^2}{4\pi}
  \frac{\sigma_\P}{\sigma_\H^2+\sigma_\P^2}-\eta_\O\ ,
\end{split}
\label{eq:diffu}
\end{equation}
In general, when we deal with protoplanetary disks, the
diffusivities $\eta_O$ and $\eta_A$ increase with larger
grains due to aforementioned analyses.

\section{Numerical Setup}
\label{sec:sim}

This letter studies PPDs that are subject to the impact of
spatial distribution of dust grains. For brevity, we refer
the readers to \citet{2018arXiv181012330W} (WBG19 hereafter)
for the details in (1) the methods of global full MHD
simulations combined with consistent thermochemistry and
ray-tracing radiation, (2) the initial conditions. For the
boundary conditions, we inherited the setups in WBG19 except
for the toroidal field above the disk region (viz. inside
the wind region) at the inner radial boundary: we set
$B_{\phi} = -B_r|_{t=0}$ (the initial value of $r$ component)
to suppress magnetic instabilities there. (note that the
magnetization here is $\sim 20$ times WBG19 in terms of
absolute magnetic pressures and stresses. The plasma $\beta$ at midplane is $2\times10^4$, 5 times smaller then WGB19) Other hydrodynamic and field components are identical to WBG19.

The dust grains are still treated as single-sized
carbonaceous grains co-moving with the gas. The size is either
$a_\gr = 5~\ang$ (Model S) or $a_\gr = 10^{-2}~\micron$
(Model L). On the other hand,
the variation of their properties is reflected by adjusting their total abundance so that the total geometric cross-section of dust grains per hydrogen nucleus, $\sigma_\gr/\chem{H}$, matches the value obtained from the dust evolution calculation (bottom panel of Figure~\ref{fig:dust_sig}), as a
function of the cylindrical radius $R$ to the central star. The basic properties of our
fiducial model (Model S) are summarized in
Table~\ref{table:fiducial-model}. Maximum
\sigperH ~is $8\times10^{-23}~\cm^2$ in both models,
which equals to $7\times10^{-6}$ in dust to gas mass ratio in
Model S, and $1.4\times10^{-4}$ in Model L, if we take the
average density of a single grain as
$2.25~\g~\cm^3$. We note that the grains in the two models have the same total
geometric cross section but couple to magnetic fields differently. The
grains in Model L have a grain-neutral collision coefficient $\langle\sigma v\rangle_\Gr$ that is two orders of magnitude larger than the grains in Model
S. The behavior of very small grains are
actually similar to ions, as the Hall parameter of ions is approximately $\beta_i\approx 3.3\times10^{-3}(B_{\rm G}/n_{15})$. For
larger charged grains, smaller $\beta_{gr}$ indicates weaker
coupling to magnetic fields, which means that grains are
more prone to frequent collision with neutrals. 

\begin{deluxetable}{lr}
  \tablecolumns{2} 
  \tabletypesize{\scriptsize}
  \tablewidth{0pt}
  \tablecaption{Properties of Model S (\S\ref{sec:sim}) 
    \label{table:fiducial-model}
  } \tablehead{ \colhead{Item} & \colhead{Value} }
  \startdata
  Radial domain & $1~\au \le r \le \ 100~\au$\\
  Latitudinal domain & $0.035\le\theta\le 3.107$ \\
  Resolution & $N_{\log r} = 384$, $N_\theta= 192$ \\
  \\
  Stellar mass & $1.0~M_\odot$ \\
  \\
  $M_\disk(1~\au \leq r \leq 100~\au)$ & $0.10~M_\odot$
  \\[2pt]
  Initial mid-plane density &
  $8.03\times 10^{14}(R/\au)^{-2.2218}~m_p~\cm^{-3}$ \\[2pt]
  Initial mid-plane plasma $\beta$ & $10^4$ \\
  Initial mid-plane temperature &
  $305(R/\au)^{-0.57}~\K$ \\
  Artificial heating profile &
  $305(R/\au)^{-0.57}~\K$ \\
  \\
  Luminosities [photon~$\s^{-1}$] & \\[5pt]
  $7~\eV$ (``soft'' FUV)  & $4.5\times 10^{42}$ \\
  $12~\eV$ (LW) & $1.6\times 10^{40}$ \\
  $3~\keV$ (X-ray) & $1.1\times 10^{38}$ \\
  \\
  Initial abundances [$n_{\chem{X}}/n_{\chem{H}}$] & \\[5pt]
  \chem{H_2} & 0.5\\
  He & 0.1\\
  \chem{H_2O} & $1.8 \times 10^{-4}$\\
  CO & $1.4 \times 10^{-4}$\\
  S  & $2.8 \times 10^{-5}$\\
  SiO & $1.7 \times 10^{-6}$\\
  \\
  Dust/PAH properties & \\
  $a_\Gr$ & $5~\ang$ \\
  $\sigma_\Gr/\chem{H}$ & Variable (\S\ref{sec:sim}) \\
  \enddata
\end{deluxetable}

\section{Results}
\label{sec:sim-results}

\begin{figure}
\centering
\includegraphics[width=0.45\textwidth, keepaspectratio]
{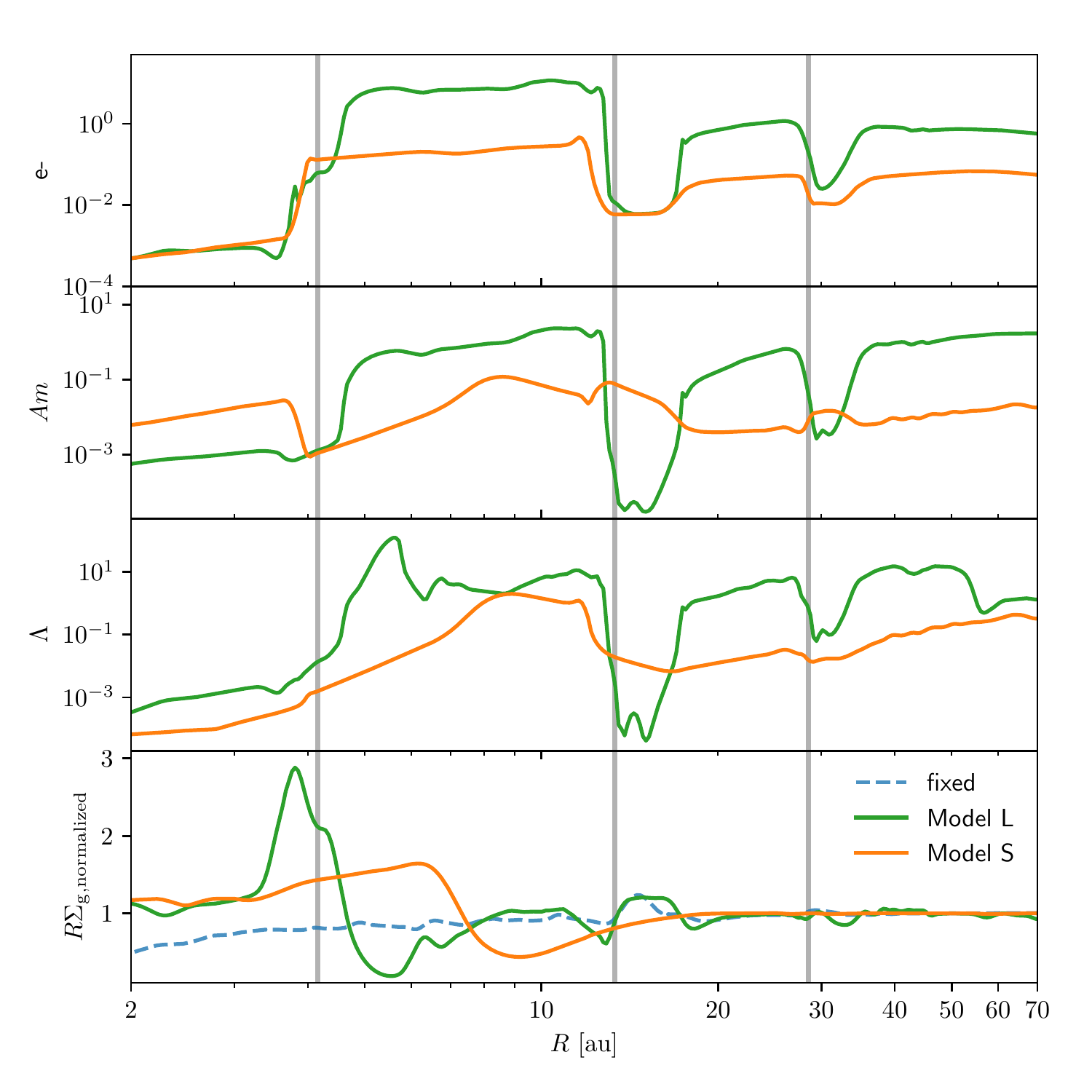}
\caption{From top to bottom: electron density
  ($\rm{cm}^{-3}$), Ambipolar Elsasser number Am, Ohmic
  Elsasser number $\Lambda$ and normalized surface
  density. Green lines are from Model L ($1\times10^{-6}$ cm
  grain) and orange lines are Model S. The blue dashed lines
  in the bottom panels (“fixed”) is the disk model with
  fixed temperature and ionization structure from
  \citet{2019arXiv190408899H}. The location of
  snow/sintering lines are marked with gray vertical lines.}
\label{fig:sigmacompmulti}
\end{figure}

Figure~\ref{fig:sigmacompmulti} summarizes the main point of
this paper: the impact of grain sizes on radial ionization
structure and how rings and gaps form from it. In Model S,
where grains have the same size as WBG19 ($5~\ang$ PAH) ,
the variation of dust number density yields some effects on
surface density beyond $10~\au$, and there are no apparent
features associated with \sigperH ~bumps at $13~\au$ and
$28~\au$. In Model L, the location of rings and gaps at
these two snow lines are very close to
\citet{2019arXiv190408899H}. From the non-ideal MHD point of
view, the simulations prove that number density variation of
PAH-scaled grains has sub-linear impacts on magnetic
diffusivities. The two middle panels shows the Elsasser
numbers of both Ohmic resistivity and Ambipolar diffusion of
Model S, which have mild change at $13~\au$ and $28~\au$ snow lines, comparing to order of magnitude jump of \sigperH. The diffusivity profiles
of Model L reflect the \sigperH ~structures quite
well. Diffusion strength increases up to 4 orders of
magnitude at those two snow lines, where the total \sigperH
~rises by about 10 times.

\begin{sidewaysfigure*}
 \centering
    \includegraphics[trim=0 0 0 -11.5cm,width=1.0\textwidth]
    {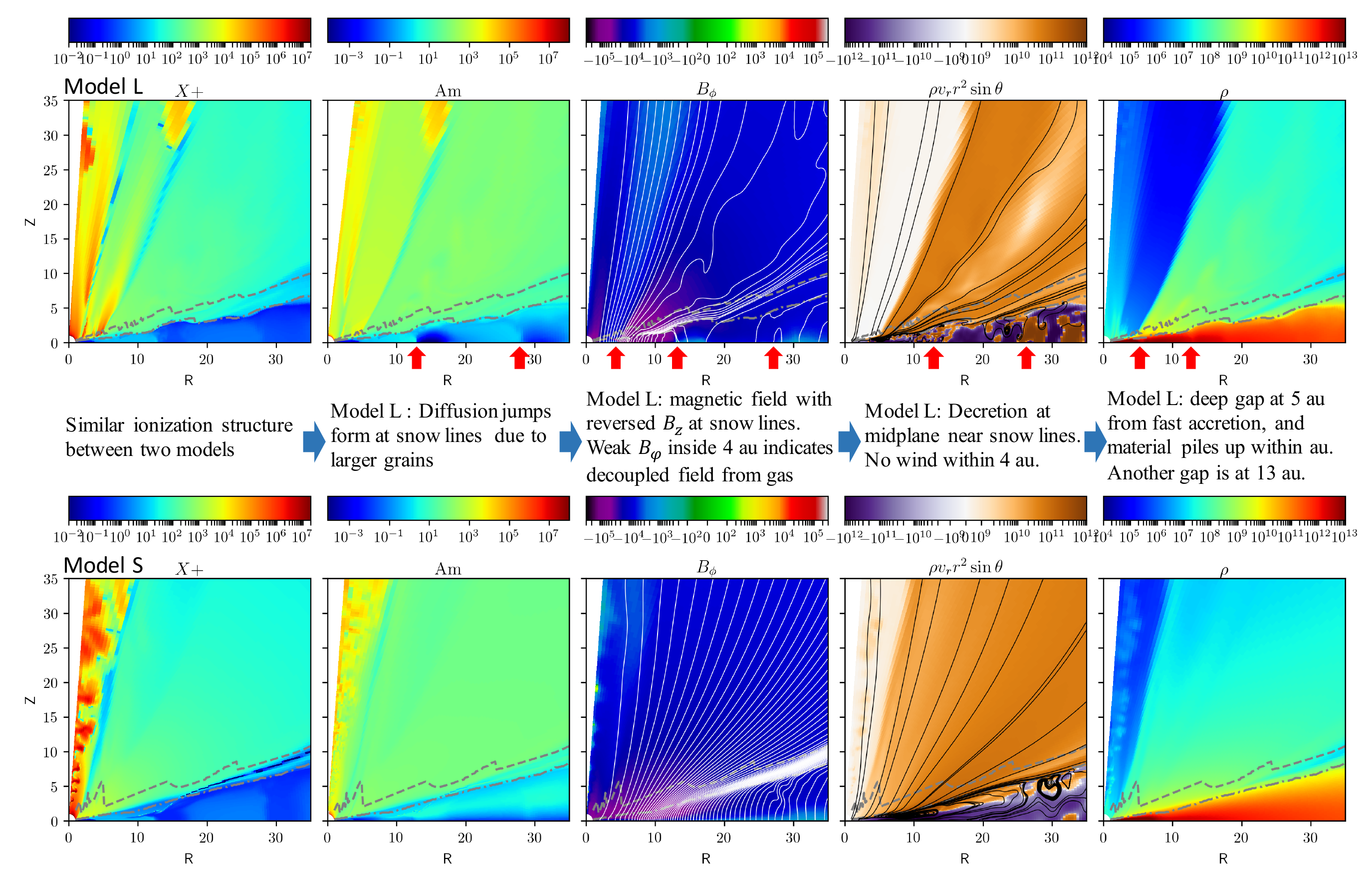}
    \caption{Meridional slices of Model L (Upper) and Model S (Lower), averaged from
      $t=1496~\yr$ to $t=1500~\yr$ with $\d t=1.0~\yr$ (except the
      \pos{X} panel is a snapshot at
      $t=1500~\yr$). From left to right we illustrate the physical process of substructure formation, especially in Model L: number density of the summation of all positive charge carriers except \pos{Gr}; Elsasser number of Ambipolar diffusion; toroidal field strength in code units, with white lines as magnetic field lines; effective mass flux and stream lines of gas velocity (black solid lines); gas density measured in number density of H atom (in $\cm^{-3}$). The gray dashed lines indicate poloidal Alfv\'en surface, and the gray dash-dotted lines are wind base/disk surface.}
\label{fig:multi2dboth}
\end{sidewaysfigure*}

After 1500 orbits at the innermost radius, the overall disk
structure of both simulations reaches a quasi-steady phase,
as shown in Figure \ref{fig:multi2dboth}. In what follows, we recognize the disk surface as the
location of wind launching, i.e., where both $v_z$ and $v_R$
are positive and poloidal velocity $\sqrt{v_z^2+v_R^2}$ is
more than 50\% of local sound speed. Above the wind base,
the Alfv\'en surface is where the velocity in the poloidal
plane equals to the poloidal Alfv\'en velocity
$v_{A,p}=\sqrt{(v_z^2+v_R^2)/4\pi\rho}$.

\subsection{Thermochemistry}
\label{sec:thermochem}
Both models exhibit vertical profiles in thermochemistry at each radius that are qualitatively similar to
\citet{2018arXiv181012330W}: a relatively warm magnetized
wind above the dense, relatively cold disk. Quantitative
properties of these vertical profiles, nonetheless, are
sensitively modulated by the radial variations in dust
effective cross section. For Model S, because of the
enhanced adsorption efficiency of charged particles at
smaller dust sizes \citet{1987ApJ...320..803D}, throughout
most radial ranges, the overall abundances of
charge-carrying species (including free electrons, ions, but not
$\chem{Gr}^\pm$ since they have little contribution to conductivity in Model L) are lower than Model L. Near the innermost
snowline of \chem{H_2O} ($R\simeq 4~\au$), Model L exhibits
significant fluctuation due to mixture of materials that
brings uneven relative abundance of dusts. Note that the co-moving dusts do not change its initial radial distribution significantly.

Because the dust-dust neutralization process
($\chem{Gr}^++\chem{Gr}^-\rightarrow 2\chem{Gr}$) is
relatively slow, charged grains ($\chem{Gr}^\pm$) become the
predominant charge carriers in the mid-plane in both models
\citep[similar to][]{2016ApJ...819...68X,
  2018arXiv181012330W}. As indicated by
Eq.~\ref{sec:grains}, charged grains behave like ions at
small sizes in terms of $\mean{\sigma v}$--whether a free
charge is carried by an ion or a charged tiny grain, it
contributes similarly to the conductivity (thus magnetic
diffusivity). Therefore, the radial variation in diffusivity
(indicated by the Elsasser numbers) in Model S is not as
intensive as the variation in dust effective cross
section. In contrast, Model L exhibits much lower Elsasser
numbers at those radii where the effective cross sections
are high, since the $a_\gr= 10^{-6}~\cm$ grains do not
contribute appreciably to the components of tensorial
conductivity due to frequent collision with neutrals (eq.\ref{eq:sig_v},\ref{eq:sigfull}). This difference induced by dust sizes are
signified by its impact on MHD diffusivities, which will be
elaborated in what follows.

\subsection{Dynamics and Kinematics of Accretions and Winds}
\label{sec:dyn-kinet}

In a (quasi-)steady accretion disk, the accretion is driven
by (1) the radial gradient of $T_{R\phi}$ (the $R\phi$
component of the magnetic stress tensor), and (2) the
difference of the $T_{z\phi}$ stress between the top and
bottom disk surfaces, namely,
\begin{equation}
  \dfrac{\dot{M}_\acc v_\K}{4\pi}
  =\dfrac{\partial}{\partial R}
  \left( R^2\int_{\rm bottom}^{\rm top} T_{R\phi}\ \d z \right)
  +R^2 \left[ T_{z\phi}\right]_{\rm bottom}^{\rm top}\ .
\label{eq:stress}
\end{equation}

The first term in eq.~\ref{eq:stress} resembles a radial
 stress and can be characterized by the equivalent
Shakura-Sunyaev $\alpha$ parameter,
\begin{equation}
  \alpha = \left[ \int_{\rm bottom}^{\rm top} p_{\rm gas}
   \ \d z\right]^{-1} \times
  \int_{\rm bottom}^{\rm top} T_{R\phi}\ \d z\,.
\label{eq:alpha}
\end{equation}
Quantities associated with these components of accretion
rates are summarized in Figure \ref{fig:mdot1dS} and
\ref{fig:mdot1dL}. The accretion rate $\dot{M}_{\textrm{acc}}$ is integrated mass
flux below the wind base, i.e., dot-dashed gray lines in Figure.\ref{fig:multi2dboth}. The wind loss rate $\dot{M}_{\textrm{wind}}$ is vertical
mass flux ($\rho v_z$) measured at wind base. Similar to \citet{2019arXiv190408899H}
and \citet{2018arXiv181012330W}, the $T_{z\phi}$ stress
(viz. the stress that drives wind) is the predominant factor
of accretion in both Models.

The second term in eq.~\ref{eq:stress} represents the
contribution of wind stresses. Along a field line, the
cylindrical radius of wind base $R_\wb$ and the
radius where it intersects with Alfven surface $R_\A$ is
related to the ratio of wind loss rate and steady accretion
rate caused by wind torque:
\begin{equation}
  \dot{M}_{\rm acc,wind}
  =2 \bigg[ \left(\frac{R_\A}{R_\wb}\right)^2-1\bigg]
  \left(\frac{\d\dot{M}_\wind}{\d\ln R}\right)\ .
\label{eq:lever}
\end{equation}
The ratio $R_\mathrm{A}/R_\mathrm{wb}$ is referred to as
``magnetic lever arm''. For example, based on the field
lines presented in the lower panel of
Figure~\ref{fig:multi2dboth}, the value of lever arm is about
$1.4$ within $10~\au$; hence, the local accretion rate
should be $1.92$ times of
$\textrm{d}\dot{M}_\textrm{wind}/\textrm{dln}R$ in Model S.

\subsubsection{Model S}

\begin{figure}
    \centering
    \includegraphics[width=0.48\textwidth]{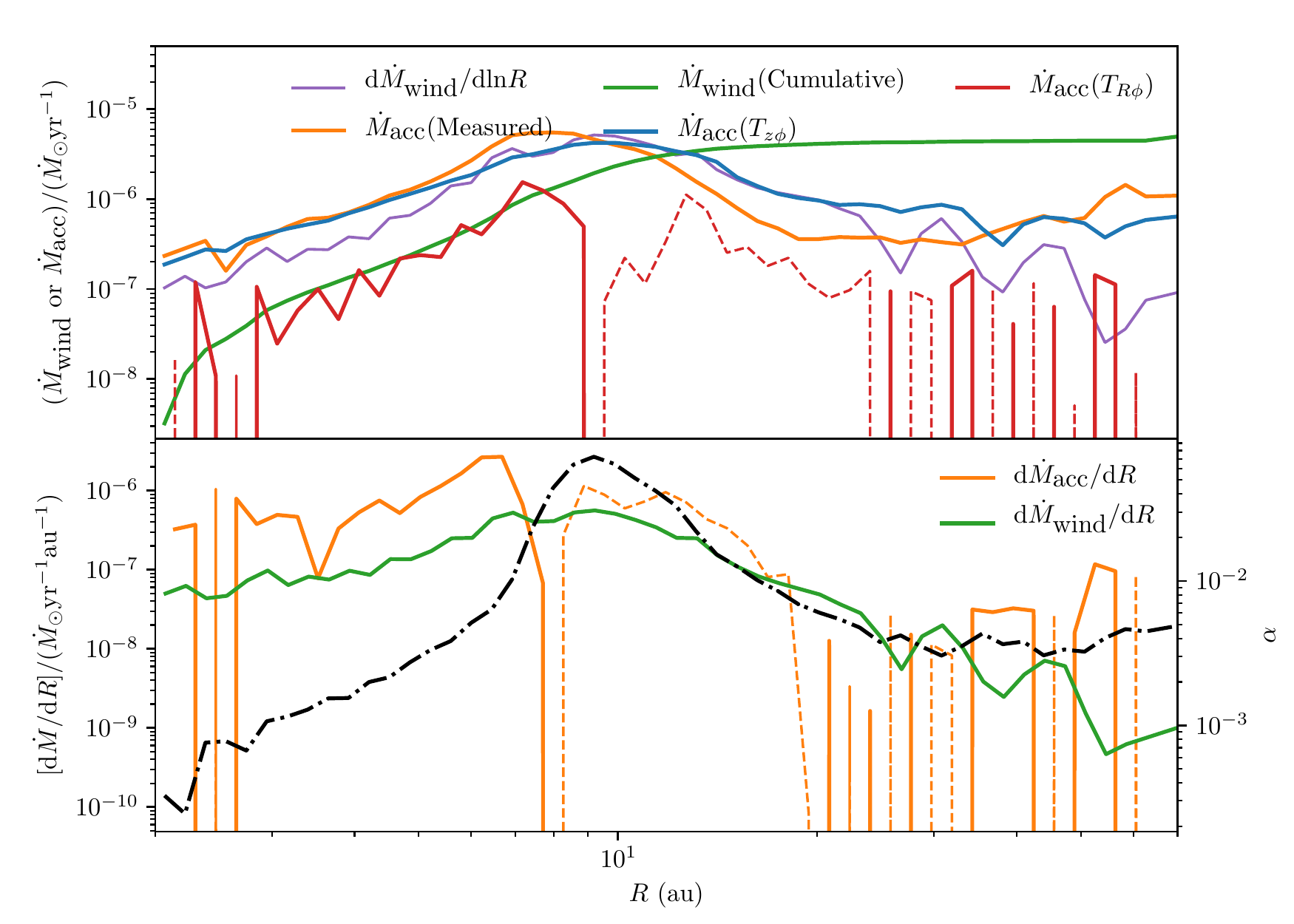}
    \caption{One-dimensional radial profiles on disk's mass
      flux, in the case of Model S. In the upper panel are
      measured accretion rate in disk ($\dot{M}_{acc}$
      (Measured), wind loss rate (both cumulative and per
      ln$R$), and estimated accretion rate caused by $R\phi$
      and $z\phi$ Maxwell stress. Lower panel shows the
      accretion rate variation by distance to star and wind
      loss rate per unit radii. The Shakura-Sunyaev $\alpha$
      parameter for radial angular momentum transport is
      also presented in black dashed-dot line on the right
      $y$ axis.}
     \label{fig:mdot1dS}
\end{figure}

For Model S, the lower panel of Figure
\ref{fig:multi2dboth} observes steady accretion below
$\sim 2/3$ the height of wind base, and outflow dominates
above that. Both velocity streamlines and magnetic field
lines in Figure \ref{fig:multi2dboth} show apparent wind
launching inside $10~\au$, while beyond $10~\au$ the
outgoing flux is almost parallel to the disk surface.  The
strengths of both poloidal and toroidal field components are
significantly weaker than $R< 15~\au$. Therefore, the wind
torque is weaker at the outer region, expecting a lower
accretion rate and slower evolution for the substructures.

Inside the disk, the Elsasser numbers $\am$ and $\Lambda$ have no apparent feature observed at the $13~\au$ and
$28~\au$ snow lines due to the effects elaborated in
\S\ref{sec:thermochem}. Radial differentiation of accretion
and wind-launching rates are most distinguishable near
$R\sim 7~\au$, at which the fields concentrate and the
accretion rate peaks
($\dot{M}_\acc\simeq 5\times10^{-6}~M_\odot~\yr^{-1}$). This profile is clearly reflected by the resulting surface
density profile (Figure~\ref{fig:sigmacompmulti}) and the
rates of wind launching (lower panel,
Figure~\ref{fig:mdot1dS}).  The accretion driven by the
$T_{z\phi}$ stress is always about one order of magnitude
higher than the contribution of the $T_{R\phi}$ stress. In
fact, the radial gradient of the $T_{R\phi}$ stress drives a layer of
{\it decretion} rather than accretion in the range
$10~\au \lesssim R \lesssim 22~\au$, indicating that the
toroidal field decreases faster than $R^{-2}$ in this radial
range.  Within $10~\au$, the accretion rate is about twice
as much as $\textrm{d}\dot{M}_\textrm{wind}/\textrm{dln}R$,
which is consistent with the magnetic lever arm argument
(eq.~\ref{eq:lever}).

\subsubsection{Model L}

\begin{figure}
    \includegraphics[width=0.48\textwidth]{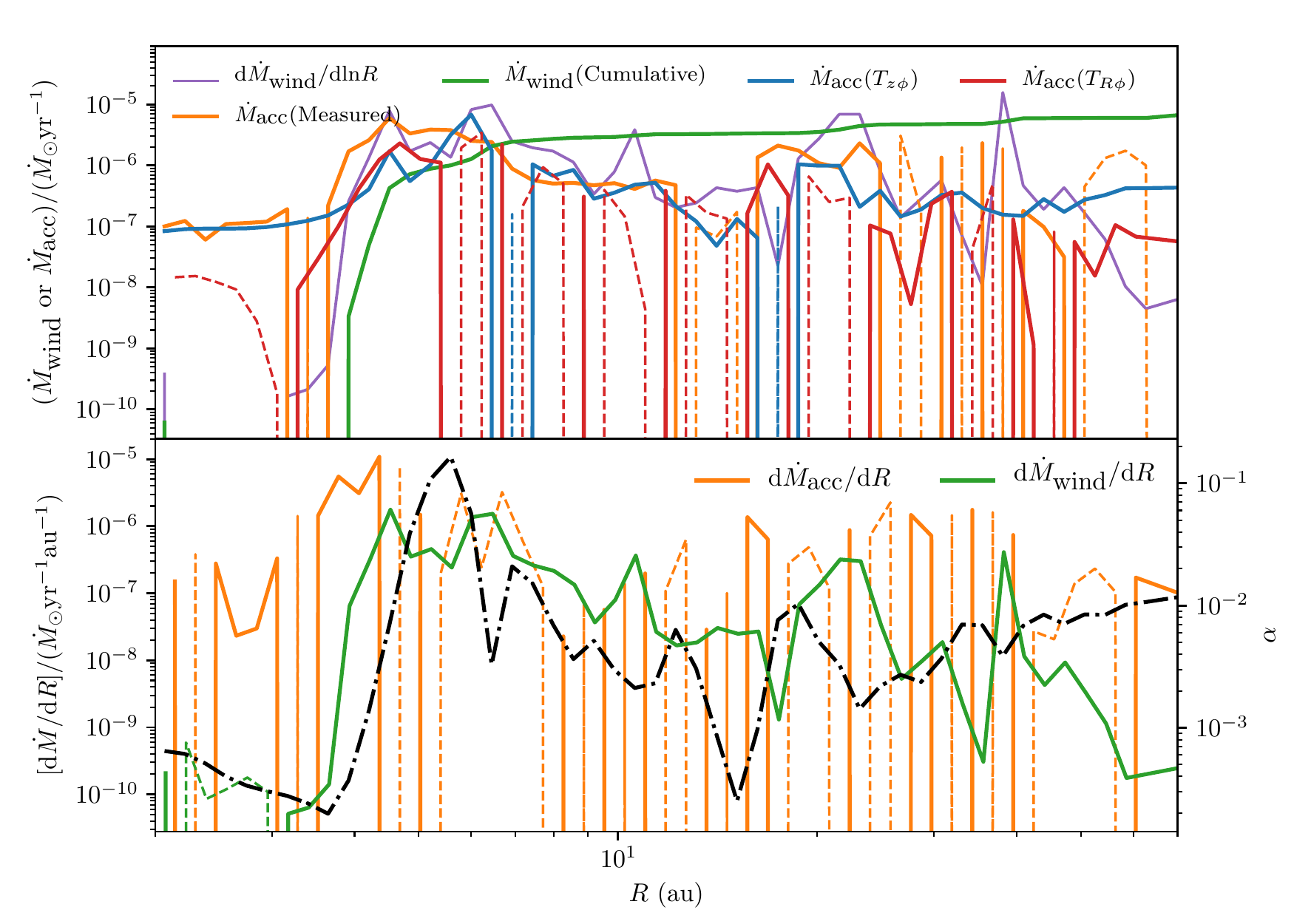}
    \caption{Same as Figure \ref{fig:mdot1dS} but for Model
      L}
    \label{fig:mdot1dL}
\end{figure}

The outcomes of our simulation are quite different in Model
L. At radii slightly above the $4~\au$, $13~\au$ and
$28~\au$ snow lines (where \sigperH ~rises significantly),
both $\am$ and $\Lambda$ decrease to yield stronger
diffusion (see also \S\ref{sec:thermochem}). Such features
in magnetic diffusion modulate the way that the disk
accretes. Figure~\ref{fig:multi2dboth} presents an obvious
characteristics of Model L: the absence of wind and
wind-driven accretion at radii $R\lesssim
4~\au$. Immediately outside this snowline radius,
concentration of magnetic fields causes excessive mass
transfer rates in both wind and accretion, causing a gap at
$R\sim 5~\au$. Qualitatively similar features examplify
themselves at snow lines at $13~\au$ and $28~\au$. Overall,
the mass transport is much more chaotic than Model S; wind
and accretion structures below the Alfv\'en surface are not
laminar. Similar to \citet{2019arXiv190408899H}, accretion and
decretion (flowing outwards but the streamlines not reaching
super-Alfv\'enic) regions are associated by poloidal field
loops at midplane, as shown in the upper middle panel of
Figure \ref{fig:multi2dboth}. From 12 to $15~\au$, the
averaged net accretion flow in the disk is even negative
(i.e. net decretion); similar decretion was reported in
\citet{2019arXiv190408899H} at approximately the same
location. Such differentiation exemplifies itself at the
$R\simeq 28~\au$ snowline. Around the local density minima
associated with these snow lines, the equivalent viscous
$\alpha$ reaches its local maxima, as the magnetic and fluid
fluxes are more turbulent there than the neighbors.
Beyond 6 \au, radial stress ($\bm{T}_\mathrm{R\phi}$)
induced accretion starts to play an equally important role
as wind ($\bm{T}_\mathrm{z\phi}$).

\subsubsection{Not to launch a wind}
One of the most distinctive feature of Model L is the absence of wind within 4 $\au$ (Figure.\ref{fig:multi2dboth}). We can understand this by analysing vertical accretion structure.
Since the mass accretion rate is
$\dot{m}_{acc}=-2\pi R\int_{\rm bot}^{\rm up}\rho v_R dz$,
we can rewrite Eq. \ref{eq:stress} in derivative form so we
can calculate accretion flux at different vertical layers:
\begin{equation}
-\rho v_R=\frac{2}{Rv_K}\frac{\partial}{\partial R}\left(R^2{\bm T}_{R\phi}\right)+\frac{2R}{v_K}\frac{\partial {\bm T}_{z\phi}}{\partial z}
\label{eq:stress2}
\end{equation}

\begin{figure}
    \centering
    \includegraphics[width=0.48\textwidth]
    {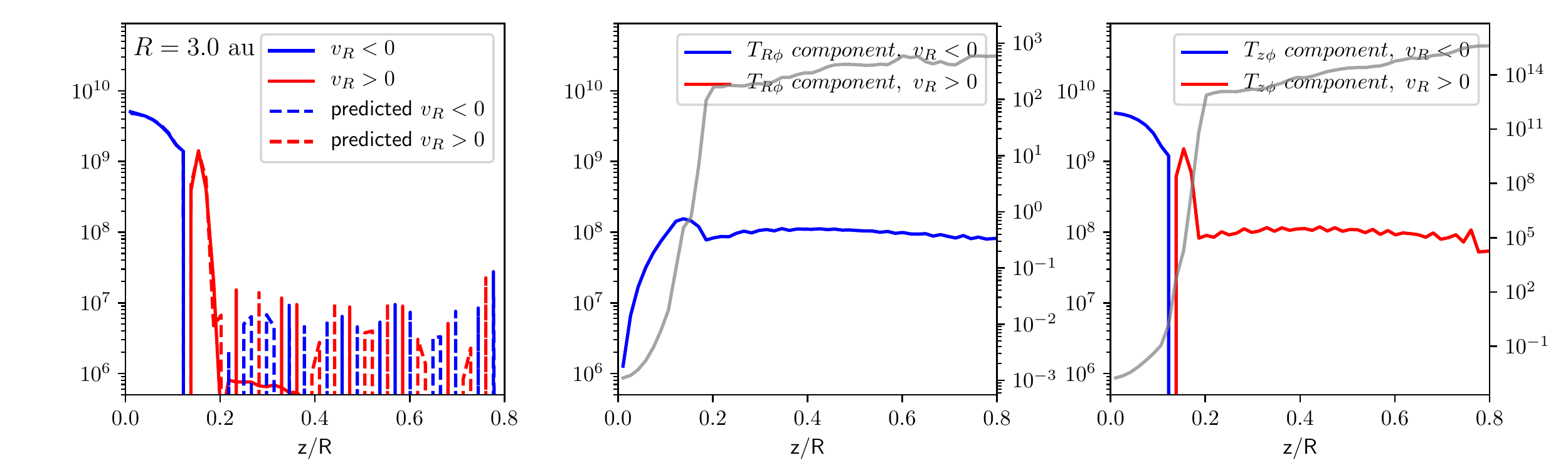}
    \includegraphics[width=0.48\textwidth]
    {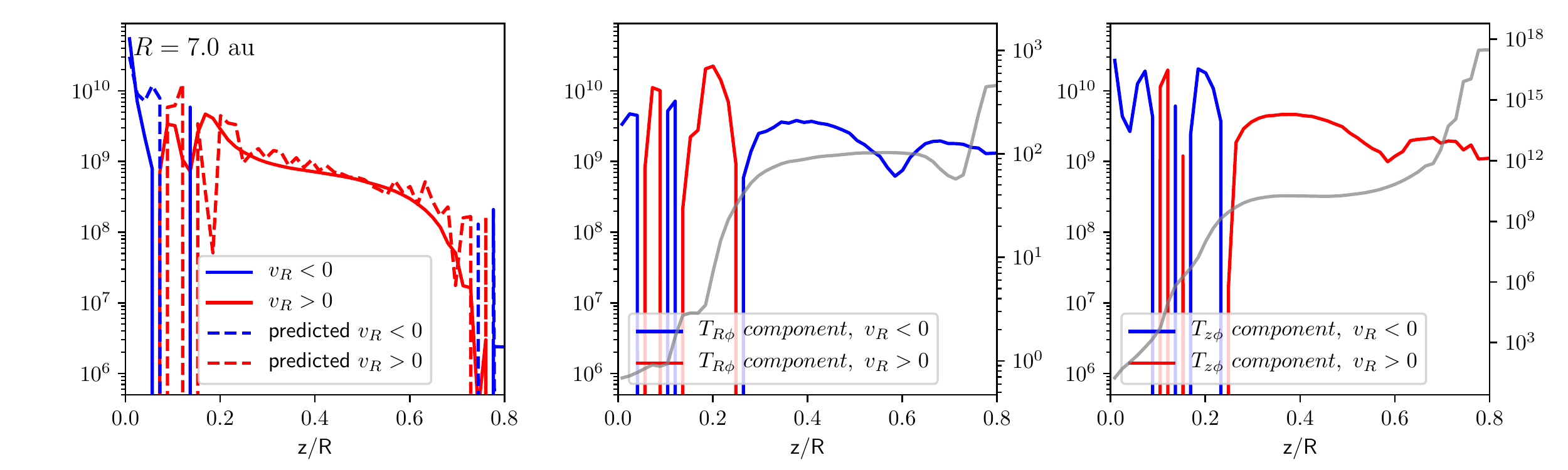}
    \caption{Vertical profiles of local mass flux [in code
      units] at R = 3 \au. Blue and red colors indicate
      accretion and decretion/outflow, respectively. In the
      left panels, solid lines are measured directly from
      the simulation, while dashed curves are values
      predicted from magnetic stress, i.e., right side of
      Eq.\ref{eq:stress2}. Middle panels show the predicted
      mass flux contributed by $R\phi$ magnetic stress, and
      the right panels show contribution from $z\phi$
      stress. The gray solid line in the middle panel shows
      same vertical profile of Ambipolar Elsasser number
      $Am$ with right side y-axis, and in the right panel is Ohmic Elsasser number
      $\Lambda$.}
    \label{fig:rhovRstress}
\end{figure}

We show vertical slice of mass flux and Maxwell stress in Model L at $R=3~\au$ and $R=7~\au$ in Figure.\ref{fig:rhovRstress}, with $R=7~\au$ being the reference. At $R=7~\au$ the disk has a "regular" structure as seen in the mass flux and density panel in Figure.\ref{fig:multi2dboth}. This is reflected in the vertical accretion analysis. Accretion is confined within a relatively thin sheet (a little more than one scale height) in the midplane, and above that we see decretion and wind. Both $\bm{T}_{R\phi}$ and $\bm{T}_{z\phi}$ drive accretion at midplane. At disk surface is the competition between $\bm{T}_{R\phi}$ driven decretion and $\bm{T}_{z\phi}$ accretion and the winner is $\bm{T}_{R\phi}$. At the wind region, $R$ component of wind is related to $\bm{T}_{z\phi}$ and it dominates over $\bm{T}_{R\phi}$ that tends to drag wind back to central star. The combination is the predicted flux at leftmost panels of Figure.\ref{fig:rhovRstress}. Mass flux at both wind region and disk region is well predicted by stress calculation.

At this point we can understand how not to launch wind at $R=3\au$. Contribution from $\bm{T}_{R\phi}$ and $\bm{T}_{z\phi}$ cancels out when $z/R>0.2$. Magnetic field lines threading the disk surface need to be bent both vertically (opening angle outward) and azimuthally (strong $\phi$ component) to lift material up. In Figure\ref{fig:multi2dboth} we find the field lines are almost vertical, respect to disk surface within $4~\au$ in Model L. 

\section{Discussion and Summary}
\label{sec:summary}

This work discusses the impact of grain sizes on the non-ideal MHD accretion processes and subsequently formation of sub-structures in PPDs. Dust grains are categorized into two populations by size, which have qualitatively different levels of impact onto the sub-structure formation in PPDs. PAH-sized grains behave as ions when charged, thus do not correspond to significant non-ideal MHD diffusivity features. Larger ($\gtrsim$sub-micron) charged grains interact with neutrals with much bigger cross-section, whose number density is related to diffusivities directly. Physical processes that affect the radial distribution of larger grains (e.g., by snow lines) lead to the formation of rings and gaps via the radial variation of accretion rates (dominating) and wind launching rates (secondary importance). 

Structure formation due to the jump of the non-ideal MHD accretion rates would lose efficiency wherever PAHs are much more abundant (in terms of effective cross-section) than sub-micron-sized grains. Therefore, the ubiquity of PPD sub-structures suggests that PAHs could be relatively rare in absence of planet-induced structure formation. This inference accords with the lack of PAH signals detected in low mass embedded YSOs \citep{2009A&A...495..837G}. In the dense regions where the temperature is relatively low, PAHs tend to freeze out and be retained by the surfaces of larger grains \citep[e.g.,][]{2017ApJ...845...13A}. 

Our non-ideal MHD simulations have connected ionization chemistry (especially dust-related processes) to sub-structure formation through accretion dynamics, but they still have a few caveats. The variation in grain sizes feeds back to the rings' locations via snow lines, as the severity of ``traffic jams'' in accretion, in turn, depends on dust size and location. Therefore, the majority of dust grains redistributed by snow lines are not necessarily the group that dominates sub-structure formation via magnetic diffusion.  In our current models, dust grains are still single-sized species co-moving with gas, which is still insufficient to account for the dust redistribution for the complete loop of feedbacks in the sub-structure formation. Radial drift alone is not efficient even for the larger grains in Model L; those grains that concentrate rapidly at pressure maxima cannot dominate the balance of ionization as they are much less in number density. Such concentration of dust, in the meantime, can also result in more fragmentation hence transfer mass in dust rings to smaller grains. Thus, a complete multi-species dust model needs to be implemented with multiple dust sizes. A viable method is to use the two-population (big and small) recipe of dust that involves coagulation and fragmentation processes for the transfer of mass between these two populations \citep[e.g.][]{2012A&A...539A.148B,2018ApJ...863...97T,2018ApJ...868...48K}. In addition, particle-based dust grain models are critical to study the vertical distribution of the two-population dusts for their sedimentation, settling, and PAH freezing processes, which are likely to concentrate the grains near the equatorial plane and make the interior of disk more susceptible to the sizes of grains.

\section*{Acknowledgements}

X.H. and Z.Z. acknowledge support from the National Aeronautics and Space Administration through the Astrophysics Theory Program with Grant No. NNX17AK40G. Simulations are carried out with the support from the Texas Advanced Computing Center (TACC) at The University of Texas at Austin through XSEDE grant TG-AST130002. L.W. acknowledges support from Center for Computational Astrophysics of Flatiron Institute. S.O. is supported by JSPS KAKENHI Grant Numbers JP16K17661, JP18H05438, and JP19K03926

\bibliographystyle{apj}

\end{CJK*}
\end{document}